\documentclass[superscriptaddress,aps,prd,twocolumn,showpacs,nofootinbib,floatfix]{revtex4-2}

\usepackage{float}
\usepackage{multirow}
\usepackage{amsmath}
\usepackage{amssymb}
\usepackage{mathtools}
\usepackage{graphicx}

\usepackage{booktabs}
\usepackage{aas_macros}
\usepackage{xcolor}
\usepackage{stackrel}
\usepackage{natbib}
\definecolor{tl}{RGB}{0,180,120}
\usepackage[colorlinks,citecolor=tl,linkcolor=red,urlcolor=magenta,unicode=true]{hyperref}

\begin{document}
\title{Constraining the Hubble parameter with the 21 cm brightness temperature signal in a universe with inhomogeneities}
\author{Subhadeep Mukherjee}   
\email{shubhadeep.avg@gmail.com}
\affiliation{Department of Astrophysics and High Energy Physics, S. N. Bose National Centre for Basic Sciences, JD Block, Sector III, Salt Lake, Kolkata-700106, India}
\author{Shashank Shekhar Pandey}
\email{shashankpandey7347@gmail.com}
\affiliation{Department of Astrophysics and High Energy Physics, S. N. Bose National Centre for Basic Sciences, JD Block, Sector III, Salt Lake, Kolkata-700106, India}
\author{A. S. Majumdar }   \email{archan@bose.res.in}
\affiliation{Department of Astrophysics and High Energy Physics, S. N. Bose National Centre for Basic Sciences, JD Block, Sector III, Salt Lake, Kolkata-700106, India}
\date{\today}

\begin{abstract}
\textbf{Abstract}
We consider the 21\,cm brightness temperature as a probe of the Hubble tension in the framework of an inhomogeneous cosmological model. Employing Buchert's averaging formalism to study the effect of inhomogeneities on
the background evolution, we consider scaling laws for the backreaction and curvature consistent with structure formation simulations. We calibrate the effective matter density using MCMC analysis using Union 2.1 Supernova Ia data. Our results show that a higher  Hubble constant ($\sim73$\,km/s/Mpc) leads to a shallower absorption feature in the brightness temperature versus redshift curve. On the other hand, a lower value ($\sim67$\,km/s/Mpc) produces a remarkable dip in the brightness temperature $T_{21}$. Such a
substantial difference is absent in the standard $\Lambda$CDM  model. Our findings indicate that inhomogeneities could significantly affect the 21\,cm signal, and may shed further light on the different measurements of the Hubble constant.

\end{abstract}
\maketitle
\section{Introduction }\label{sec:intro}

Although the standard cosmological paradigm described by the $\Lambda$CDM 
model has successfully explained a wide range of cosmological phenomena, it currently faces several notable challenges \cite{2025_LDCDM_tensions,colgáin2025_1}. For instance, significant tensions in the measurement of the Hubble constant have emerged, with local determinations yielding higher values than those inferred from cosmic microwave background observations \cite{Riess_2021, Riess_2022, 2020}. Locally, measurements using distance ladder methods employing standard candles such as Cepheid variables and Type Ia supernovae yield a higher $H_0$, around $73$ km/s/Mpc \cite{Riess_2022}. In contrast, analysis of the cosmic microwave background (CMB) data from Planck \cite{Planck_2015, Planck_2018}, which models the universe's evolution from recombination under the $\Lambda$CDM framework, suggests a lower $H_0$, around $67$ km/s/Mpc \cite{Planck_2018}. This discrepancy \cite{Dainotti_2021,Dainotti_2022,dainotti_2025}, well beyond the expected error margins, indicates unidentified systematic errors in one or both methods, or the need for new physics beyond the standard model.

Additionally, recent James Webb Space Telescope observations have revealed massive high-redshift galaxies that further strain the model \cite{Labb__2023, Boylan_Kolchin_2023}. Moreover, observations of large-scale structures indicate pronounced inhomogeneities in the matter distribution. Analysis of galaxy distributions from the Two Degree Field Galaxy Redshift Survey reveals large amplitude fluctuations extending up to the survey boundaries, suggesting the presence of structures exceeding 100 Mpc/$h$ \cite{Labini_2009}. Similarly, studies of Luminous Red Galaxy samples from the Sloan Digital Sky Survey have reported statistically significant deviations—exceeding 2$\sigma$—from $\Lambda$CDM mock catalogues on scales as large as 500 $h^{-1}$ Mpc \cite{wiegand_scale}. Furthermore, a colossal arc of galaxies covering a distance of nearly one gigaparsec (Gpc) has been discovered \cite{lopez}, presenting further challenges to the standard model scenario, although its cosmological interpretation has been questioned in later works \cite{lopez2025,sawala2025,sawala2025_1}.

Given the above observations and challenges, multi-messenger approaches are increasingly important: combining electromagnetic, gravitational-wave and neutrino observations can yield a more complete picture of cosmic history. In particular, the $21$ cm line of neutral hydrogen observed in the radio band provides a powerful and complementary electromagnetic probe of the cosmic dawn and the epoch of reionisation. The 21 cm line, a hyperfine transition of the neutral hydrogen atom, corresponding to a wavelength of 21 cm or a frequency of $1420$ MHz, arises from the energy difference between the parallel and antiparallel spin states of the electron and proton in the neutral hydrogen atom, the most abundant element which constitutes almost 75 $\%$ of the total baryonic matter content of the universe. It captures the thermal and ionization state of the universe, which makes it an essential tool for studying cosmological phenomena \cite{BARKANA2001, FURLANETTO2006181, Pritchard_2012, BARKANA2016}, ranging from the epoch of re-ionization (EoR) \cite{Madau_1997} to the formation of the first stars \cite{Fazio:421211,10.1088/2514-3433/ab4a73}. The intensity of the redshifted 21 cm signal from neutral hydrogen is commonly quantified using the brightness temperature $T_{21}$, which measures the difference between the spin temperature of hydrogen gas and the cosmic microwave background (CMB) temperature. The brightness temperature depends on factors such as the neutral hydrogen fraction, gas density, and interactions with radiation fields, making it a key observable in 21 cm cosmology \cite{FURLANETTO2006181, Pritchard_2012}.

\par The EDGES collaboration \cite{EDGES_Bowman2018} drew significant attention with their findings on the $T_{21}$ signal within the redshift range of $15 \leq z \leq 20$, reporting a $T_{21}$ value of $-500^{+200}_{-500}$ mK. This measurement exceeds the expectations from the standard $\Lambda$CDM cosmological model by over twofold. Though a later experiment, SARAS \cite{SARAS_Singh2022} has found no evidence of such a signal, considerable current effort
is directed towards the search and analysis of the 21 cm signal due to its potential as a key probe in multi-messenger astronomy. Several possibilities of generating the excess $T_{21}$ signal have
been investigated, such as cooling mechanisms based on  models of dark energy \cite{halder_global_2022}, and baryon-dark matter scattering \cite{munoz_haimound, Halder_2021, Datta_PRD, Ashadul_mnras}, as well as by considering an excess radio background above CMB \cite{Feng_2018, Ewall, Fialkov, Ewall2}. Furthermore, studies have utilized the $T_{21}$ signal as a tool to place constraints on the abundance of primordial black holes (PBHs) \cite{Clark_2018,yang2020,Mittal_2022}.

\par In the present study, we make use of a cosmological model characterized by an inhomogeneous distribution of matter, which could extend up to just below the Hubble scale, in order to examine the brightness temperature associated with the 21 cm signal. The approach used offers an alternative perspective to the standard $\Lambda$CDM framework, which assumes a statistically homogeneous and isotropic universe on large scales - a premise that has been subject to ongoing scrutiny \cite{Kumar_Aluri_2023}. Taking into account the collective inconsistencies noted in observational data, along with the characterization of the $\Lambda$CDM model as primarily a phenomenological framework tailored to align with empirical observations, rather than being anchored in fundamental theoretical principles \cite{ellis2006, Clarkson_2010}, it becomes apparent that investigating alternative cosmological models could potentially provide a similarly robust, if not more comprehensive, portrayal of cosmic phenomena.
  
\par To integrate the concept of inhomogeneity into the framework, it is essential to employ an appropriate averaging process. The literature provides a variety of proposed averaging methodologies, as documented in several works \cite{Ellis1984, Futamase, Zalaletdinov1992, Zalaletdinov1993, Gasperini_2011}. In our analysis, we specifically select the Buchert averaging technique \cite{Buchert, Buchert2001}. The choice of the Buchert averaging approach is grounded in its unique capability to streamline the averaging of scalar quantities, translating them into observationally measurable parameters, such as the relationships between redshift and angular diameter distances \cite{rasanen1, rasanen2, Koksbang_2019, Koksbang2, Koksbang3, Koksbang4}. Numerous investigations have deployed the Buchert averaging formula to explore the role of inhomogeneities within cosmological dynamics. Such studies often aim to elucidate the characteristics of the universe without invoking exotic physics \cite{Coley, Buchert, Buchert2001, Korzynski_2010, Clifton, Skarke, Buchert_2015, Buchert4, Weigand_et_al, Rasanen_2004, wiltshire, Kolb_2006, Koksbang_2019, Koksbang2, Koksbang_PRL, Rasanen_2008, bose, Bose2013, Ali_2017, Pandey_2022, Pandey2023, Koksbang5, Koksbang6, Koksbang7, Koksbang8, Halder_2023, Koksbang2023, T21_shas}.

\par For demonstration of our approach, in this work we consider a simple two-domain model \cite{Wiltshire_2007,wiltshire_2007_1}, wherein the global domain $\mathcal{D}$ is partitioned into regions denoted as overdense $\mathcal{M}$ and underdense $\mathcal{E}$. Our study investigates the global 21 cm brightness temperature signal in this inhomogeneous universe setting, excluding any exotic forms of cooling, besides the standard $\mathrm{Ly\alpha}$ effect \cite{Meiksin, Meiksin20,tstb1, Chuzhoy_2007, Ghara, Mittal_2020}. We calibrate our model parameters by performing Markov Chain Monte Carlo (MCMC) analysis \cite{mcmc1,mcmc2} with Union 2.1 Supernova Ia data \cite{union} to determine the optimal value of our model parameters using structure formation history derived from N-body Newtonian simulation data \cite{simulation}. We explore the potential of 21 cm brightness temperature as a tool to probe the Hubble tension scenario. Using the best-fit values obtained for our model parameters using MCMC analysis, we compute the brightness temperature for both sets of $H_0$ values. Notably, our results show that the signal exhibits a more pronounced temperature dip for the $H_0$ value aligned with Planck's observations. This result indicates that incorporating inhomogeneous averaging could offer a fresh perspective in reconciling the differing $H_0$ estimates and thus shed light on the Hubble tension problem.

\par The structure of the paper is organized as follows. First, we briefly introduce the methodology for evaluating the brightness temperature of the 21 cm signal in \autoref{sec:21cm}. Following this, we present an overview of Buchert's averaging technique in \autoref{sec:buchert}. Next, in \autoref{sec:scale}, we introduce a new scaling solution that incorporates the effects of structure formation. The analysis of the 21 cm signal from our two-domain model is discussed in \autoref{sec:par_plot}. In \autoref{sec:ob}, we employ the Union 2.1 supernova Ia data set to constrain our model parameters and determine the 21 cm brightness temperature for the redshift range $15<z<30$, using the optimal parameter values. We summarize our findings in \autoref{sec:result}.

\section{\label{prelim} Preliminaries}

\subsection{\label{sec:21cm} 21 cm Brightness Temperature}

The expression for brightness temperature for the 21 cm signal can be written as \cite{Ashadul_PRD, munoz_21cm},

\begin{equation}\label{eq:T21}
    T_{21} = \frac{T_s - T_\gamma}{1+z}(1-e^{-\tau(z)}),
\end{equation}
where $T_s$ and $T_\gamma$ denote the spin and cosmic microwave background (CMB) temperatures, respectively.
The optical depth $\tau(z)$ is defined as \cite{munoz_21cm},

\begin{equation}\label{eq:tau}
    \tau(z) = \frac{3}{32\pi}\frac{T_*}{T_s}n_{\mathrm{HI}}\lambda_{21}^3\frac{A_{10}}{H(z)+(1+z)\delta_r v_r},
\end{equation}
where $T_* = \frac{hc}{k_B\lambda_{21}} \approx 0.068\ \mathrm{K}$, $A_{10} = 2.85\times 10^{-15}\ \mathrm{s^{-1}}$, is the Einstein A-coefficient \cite{Haimoud_Hirata_2010}, $\lambda_{21} \approx 21\ \mathrm{cm}$, $n_{\mathrm{HI}}$ is the neutral hydrogen density, $H_(z)$ is the Hubble parameter which depends on the specific cosmological model we choose, and $\delta_r v_r$ is the gradient of peculiar velocity along the line of sight and is only significant at local regions. $T_{21} < 0$ signifies the absorption characteristic and $T_{21} > 0$ signifies the emission characteristic.

The spin temperature $T_s$ governs the ratio of the population of the upper and lower hyperfine levels of the H-atom and is given by \cite{munoz_21cm} as,

\begin{equation}\label{eq:Ts1}
    \frac{n_1}{n_0} = 3 e^{-T_*/T_s},
\end{equation}
where $n_1$ and $n_0$ denote the number densities of hydrogen atoms in their excited and ground states, respectively.
The spin temperature $T_s$ is determined by the overall interaction of the neutral hydrogen atom present in the intergalactic medium (IGM) with CMB photons, interaction with $\mathrm{Ly\alpha}$ photons, and collision with other hydrogen atoms/electrons. Mathematically, we calculate $T_s$ as \cite{Zaldarriaga_2004, Clark_2018},

\begin{equation}\label{eq:Ts2}
    T_s = \frac{T_\gamma + y_c T_b + y_{\mathrm{Ly\alpha}}T_{\mathrm{Ly\alpha}}}{1 + y_c + y_{\mathrm{Ly\alpha}}},
\end{equation}
where $y_c$ is  the collisional coupling coefficient, $T_b$ denotes the baryon temperature, and $y_{\mathrm{Ly\alpha}}$ is  the $\mathrm{Ly\alpha}$ coupling coefficient which corresponds to the Wouthuysen-Field effect \cite{Wouthuysen, Field}. The quantity $T_{\mathrm{Ly\alpha}}$ denotes the $\mathrm{Ly\alpha}$ background temperature. 

From the equation of $T_s$ (\autoref{eq:Ts2}), we can see that the spin temperature $T_s$ is essentially the weighted arithmetic mean of the CMB temperature $T_{\gamma}$, the baryon temperature $T_b$ and the $\mathrm{Ly\alpha}$ temperature $T_{\mathrm{Ly\alpha}}$. 
The coefficients $y_c$ and $y_{\mathrm{Ly\alpha}}$ are defined as $y_c = \frac{C_{10}T_*}{A_{10}T_b}$ and $y_{\mathrm{Ly\alpha}} = \frac{P_{10}T_*}{A_{10}T_{\mathrm{Ly\alpha}}}$ \cite{Kuhlen_2006}, respectively, with $C_{10}$ and $P_{10}$ being the coupling de-excitation rates due to collisions and $\mathrm{Ly\alpha}$ photons, respectively. 
We have $P_{10} \approx S_\alpha J_{\alpha}$, with $S_\alpha$ \cite{Hirata_2006} and $J_{\alpha}$ being the spectral distortion and the $\mathrm{Ly\alpha}$ background intensity, respectively. The values of $J_\alpha$ and $S_\alpha$ can be estimated following \cite{Mittal_2020, Clark_2018, Ciardi_2003}.

\par For the thermal evolution equation of $T_b$, we do not consider any exotic cooling or heating effect. Here we only consider the heating caused by $\mathrm{Ly\alpha}$ photons.  $T_b$ is obtained by solving the thermal equation \cite{Mittal_2020}, 

\begin{equation}
    (1+z)\frac{dT_b}{dz} = 2T_b - \frac{8\pi}{3}\frac{\mathfrak{h}}{k_B\lambda_\alpha}\frac{J_\alpha(z)\Delta\nu_D}{n_b(z)}(I_c + rI_i), \label{eq:Tb}
\end{equation}
where $\mathfrak{h}$ is the Planck's constant, $k_B$ is Boltzmann constant, $\lambda_\alpha$ is the wavelength of $\mathrm{Ly\alpha}$ photons, $n_b = n_H(1 +x_{He} + x_e)$ is the total particle number density with $x_{He}$ and $x_e$ being the helium and electron fraction, respectively. We can directly calculate $x_{He}$ using the relation $x_{He} = \frac{Y_p}{4(1 - Y_p)}$, where $Y_p$ is the primordial He abundance, but for $x_e$ one needs to solve the thermal evolution \cite{haimoud_hirata_2011, madhava_21cm}, 

\begin{equation}
    (1+z)\frac{dx_e}{dz} = \frac{C_P}{H(z)}\left(n_H\alpha_Bx_e^2-4(1-x_e)\beta_Be^{-\frac{3E_0}{4k_BT_\gamma}}\right), \label{eq:xe}
\end{equation}
where $n_H$ is the total number density of hydrogen atoms and $E_0$ represents the ionization energy of the Hydrogen atom. The Peebles $C$ factor, $C_P$, is given by \cite{Peebles, haimoud_hirata_2011},

\begin{equation}\label{eq:C_P}
    C_P = \frac{\frac{3}{4}R_{\mathrm{Ly\alpha}} + \frac{1}{4}\Lambda_{2s1s}}{\beta_B + \frac{3}{4}R_{\mathrm{Ly\alpha}} + \frac{1}{4}\Lambda_{2s1s}},
\end{equation}
where $R_{\mathrm{Ly\alpha}} = \frac{8\pi {H(z)}}{3n_H(1-x_e)\lambda^3_{\mathrm{Ly\alpha}}}$ is the $\mathrm{Ly\alpha}$ photon escape rate, with $\Lambda_{2s,1s} \approx 8.22\ \mathrm{s^{-1}}$ \cite{haimoud_hirata_2011}. The recombination and ionization coefficients, $\alpha_B$ and $\beta_B$, can be calculated as described in \cite{Ashadul_mnras, Ashadul_PRD}.
 The remaining terms $\Delta\nu_D$ , $I_c$ , $I_i$ and $r$ of (\autoref{eq:Tb}) are defined by \cite{Mittal_2020,FP06},
 
\begin{widetext}

\begin{eqnarray}
\Delta\nu_D = \nu_\alpha\sqrt{\frac{2k_BT_b}{m_Hc^2}}\\
I_c = \eta\sqrt[3]{2\pi^4a^2\tau^2_\alpha} [A^2_i(-\xi_2) + B^2_i(-\xi_2)] , \text{where} \hspace{0.1 cm} \xi_2 = \eta\sqrt[3]{\frac{4\alpha \tau_\alpha}{\pi}}\label{I_c}\\
I_i = \eta\sqrt\frac{a \tau_\alpha}{2}\int_{0}^{\infty}\exp{\left[-2\eta y - \frac{\pi y^3}{6a\tau_\alpha}\right]} \text{erfc}\sqrt{\frac{\pi y^3}{2a \tau_\alpha}}\,\frac{dy}{\sqrt{y}} \ -\frac{S_{\alpha}(1-S_{\alpha})}{2\eta}\label{I_i}\\
r = \frac{J^i_\alpha}{J^c_\alpha}\label{r_frac}
\end{eqnarray}
\end{widetext}

where $y=\frac{\nu_{\alpha}-\nu}{\Delta \nu_{D}}$, $\eta = \frac{h/\lambda_\alpha}{\sqrt{2m_Hk_BT_b}}$, $a = \frac{A_\alpha}{4\pi\Delta\nu_D}$ and $\tau_\alpha = \frac{3\gamma_\alpha\lambda_\alpha^3n_Hx_{HI}}{2H}$ \cite{GP} are dimensionless frequency variable, recoil parameter, Voigt parameter and $\mathrm{Ly\alpha}$ optical depth respectively. Here $A_\alpha$ is the Einstein spontaneous emission coefficient, $\gamma_\alpha = 50$MHz \cite{Hirata} is the half width at half maximum of $\mathrm{Ly\alpha}$ resonance line. $\Delta\nu_D$ is called the Doppler width, $\lambda_\alpha(\nu_\alpha)$ is the wavelength(frequency) of the $\mathrm{Ly\alpha}$ photon, $m_H$ is the mass of the recoiling hydrogen atom and $x_{HI} \approx 1 - x_e$ is the fraction of neutral hydrogen atom which is approximately equal to 1 in the redshift range $14\leq z \leq 30$. 

\par The terms $I_c$ and $I_i$ are  related to the heating and cooling effects by $\mathrm{Ly\alpha}$ photons. Specifically speaking, the term $I_c$ is related to heating due to $\mathrm{Ly\alpha}$ continuum photons, and $I_i$ gives the cooling effect due to $\mathrm{Ly\alpha}$ injected photons except at extremely low temperature $(T \leq 1 K)$. Finally the term $r$ in (\autoref{r_frac}) which is defined as fraction of background $\mathrm{Ly\alpha}$ intensity for injected photons $J^i_\alpha$ to that of continuum photons $J^c_\alpha$ is to compensate for the fact that $I_c$ and $I_i$ in (\autoref{I_c}) and (\autoref{I_i}) are evaluated by assuming the background $\mathrm{Ly\alpha}$ intensity to be the same for both continuum $(I_c)$ and injected $(I_i)$ $\mathrm{Ly\alpha}$ photons which is generally not the case. We take $r = 0.1$ appropriate for Population-II (Pop-II) stars \cite{Mittal_2020, Chuzhoy_2007}.

\subsection{Buchert's backreaction formalism}\label{sec:buchert}

In this analysis, we employ Buchert's averaging method for a pressure-less (dust universe) model \cite{Buchert, buchert_rasanen}. Buchert's backreaction formalism reduces the complexity of the averaging problem by focusing solely on scalar quantities. The spacetime is partitioned into flow-orthogonal hypersurfaces, described by the line element \cite{Buchert, Weigand_et_al}

\begin{equation}\label{eq:line_element}
    ds^2 = -dt^2+g_{ij}dX^idX^j,
\end{equation}
with $t$ representing the proper time, $X^i$ denoting Gaussian normal coordinates on the hypersurfaces, and $g_{ij}$ being the spatial three-metric corresponding to hypersurfaces having constant $t$. The volume of a compact spatial region $\mathcal{D}$ on these hypersurfaces is described as,

\begin{equation}\label{eq:volD}
    |\mathcal{D}|_g := \int_{\mathcal{D}} d\mu_g
\end{equation}
where $d\mu_g:= \sqrt{\prescript{(3)}{}{g(t,X^1,X^2,X^3)}}dX^1dX^2dX^3$.  A dimensionless (`effective') scale factor is defined as

\begin{equation}\label{eq:scale_factor}
    a_{\mathcal{D}}(t) := \left(\frac{|\mathcal{D}|_g}{|\mathcal{D}_i|_g}\right)^{1/3},
\end{equation}
normalized with respect to a certain volume which we take as the present volume of the original domain $|\mathcal{D}_i |_g$,  representing the domain volume at the present time $t_0$. The mean value of a scalar quantity $f$ is given by,

\begin{equation}\label{eq:average_scalar}
    \langle f\rangle_{\mathcal{D}}(t) := \frac{\int_{\mathcal{D}}f(t,X^1,X^2,X^3)d\mu_g}{\int_{\mathcal{D}}d\mu_g}
\end{equation}

Employing Buchert’s scalar averaging technique for a universe consisting of pressureless dust, we average the scalar parts of the Einstein equations, that is, the Hamiltonian constraint and the Raychaudhuri evolution equation for the expansion scalar, together with the continuity equation. This is done under the assumptions of flow-orthogonal hypersurfaces, zero vorticity, and a spatially compact averaging domain, yielding the following evolution equations:

\begin{eqnarray}\label{eq:aD_ddot}
    3\frac{\ddot{a_{\mathcal{D}}}}{a_{\mathcal{D}}} = -4\pi G\langle\rho\rangle_{\mathcal{D}} + \mathcal{Q}_{\mathcal{D}} + \Lambda\\
    3H^2_{\mathcal{D}} = 8\pi G\langle\rho\rangle_{\mathcal{D}} - \frac{1}{2}\langle\mathcal{R}\rangle_{\mathcal{D}} - \frac{1}{2}\mathcal{Q}_{\mathcal{D}} + \Lambda\label{eq:aD_dot}\\
    0 = \partial_t\langle\rho\rangle_{\mathcal{D}} + 3H_{\mathcal{D}}\langle\rho\rangle_{\mathcal{D}}\label{eq:continuity}
\end{eqnarray}
where $\langle\rho\rangle_{\mathcal{D}}$, $\langle\mathcal{R}\rangle_{\mathcal{D}}$, and $H_{\mathcal{D}}$ represent the domain $\mathcal{D} $'s averaged matter density, averaged spatial Ricci scalar, and Hubble parameter (defined as $H_{\mathcal{D}}:= \dot{a_{\mathcal{D}}}/a_{\mathcal{D}}$), respectively.           $\mathcal{Q}_{\mathcal{D}}$ is called the kinematical backreaction and is defined as 

\begin{equation}\label{eq:Q_D}
    \mathcal{Q}_{\mathcal{D}}:= \frac{2}{3}(\langle\theta^2\rangle_{\mathcal{D}} - \langle\theta\rangle^2_{\mathcal{D}}) - 2\langle\sigma^2\rangle_{\mathcal{D}},
\end{equation}
where $\theta$ represents the local expansion rate and $\sigma^2 := \frac{1}{2} \sigma_{ij}\sigma^{ij}$ denotes the squared rate of shear. The Hubble parameter $H_{\mathcal{D}}$ and the averaged expansion rate $\langle\theta\rangle_{\mathcal{D}}$ are connected through the equation $H_{\mathcal{D}} = \frac{1}{3}\langle\theta\rangle_{\mathcal{D}}$. For a domain similar to an FLRW model, $\mathcal{Q}_{\mathcal{D}}$ equals zero. The integrability condition that links (\autoref{eq:aD_ddot}) and (\autoref{eq:aD_dot}) is given by,

\begin{equation}\label{eq:integrability}
    \frac{1}{a^2_{\mathcal{D}}}\partial_t(a^2_{\mathcal{D}}\langle\mathcal{R}\rangle_{\mathcal{D}}) + \frac{1}{a^6_{\mathcal{D}}}\partial_t(a^6_{\mathcal{D}}\mathcal{Q}_{\mathcal{D}}) = 0.
\end{equation}
(\autoref{eq:integrability}) illustrates a key aspect of the averaged equations, highlighting the interplay between the averaged intrinsic curvature $(\langle\mathcal{R}\rangle_{\mathcal{D}})$ and the kinematical backreaction term $(\mathcal{Q}_{\mathcal{D}})$, which represents the effect of matter inhomogeneities. This connection between $\langle\mathcal{R}\rangle_{\mathcal{D}}$ and $\mathcal{Q}_{\mathcal{D}}$, along with the term $\mathcal{Q}_{\mathcal{D}}$, indicates the deviation from homogeneity.

Within the Buchert formalism, we now adopt a specific method in which the global domain is represented by ensembles of disjoint regions \cite{Buchert_2015, Buchert4, Weigand_et_al, Rasanen_2004, wiltshire, Kolb_2006, Koksbang_2019, Koksbang2, Koksbang_PRL, Rasanen_2008, bose, Bose2013, Ali_2017, Pandey_2022, Pandey2023, Koksbang5, Koksbang6, Koksbang7, Koksbang8}. In this context, the global domain $\mathcal{D}$ is partitioned into subregions $\mathcal{F}_l$, each comprising distinct elementary spatial entities $\mathcal{F}_l^{(\alpha)}$. Mathematically, this is expressed as $\mathcal{D} = \cup_l\mathcal{F}_l$, where $\mathcal{F}_l:=\cup_{\alpha}\mathcal{F}^{(\alpha)}_l$ and $\mathcal{F}_l^{(\alpha)}\cap \mathcal{F}_m^{(\beta)} = \emptyset$ for all $\alpha\neq\beta$ and $l\neq m$.

The averaged value of a scalar function $f$ over the domain $\mathcal{D}$ is defined as,

\begin{equation}\label{eq:averaging}
\begin{split}
    \langle f\rangle :&= |\mathcal{D}|^{-1}_g \int_\mathcal{D} fd\mu_g
    = \sum_l |\mathcal{D}|_g^{-1}\sum_{\alpha}\int_{\mathcal{F}_l^{(\alpha)}} f d\mu_g\\
    &= \sum_l\frac{|\mathcal{F}_l|_g}{|\mathcal{D}|_g}\langle f\rangle_{\mathcal{F}_l} = \sum_l \lambda_l\langle f\rangle_{\mathcal{F}_l}
\end{split}
\end{equation}

where 

\begin{equation}\label{eq:lambda}
    \lambda_l := \frac{|\mathcal{F}_l|_g}{|\mathcal{D}|_g}
\end{equation}
is the volume fraction of the subregion $\mathcal{F}_l$ such that $\sum_l\lambda_l = 1$ and $\langle f\rangle_{\mathcal{F}_l}$ represents the average of $f$ within the subregion $\mathcal{F}_l$. The scalar quantities $\rho$, $\mathcal{R}$, and $H_\mathcal{D}$ are described by (\autoref{eq:averaging}). However, $\mathcal{Q}_{\mathcal{D}}$, due to the presence of the $\langle\theta\rangle_{\mathcal{D}}^2$ term, obeys

\begin{equation}\label{eq:QDsum}
    \mathcal{Q}_{\mathcal{D}} = \sum_l\lambda_l\mathcal{Q}_l + 3\sum_{l\neq m}\lambda_l\lambda_m(H_l-H_m)^2.
\end{equation}
In the subregions $\mathcal{F}_l$, $\mathcal{Q}_l$ and $H_l$ are defined analogously to how $\mathcal{Q}_{\mathcal{D}}$ and $H_{\mathcal{D}}$ are defined within the domain $\mathcal{D}$.

The scale factor $a_l$ for a subregion $\mathcal{F}_l$ can also be defined. Since the subregions are disjoint by definition, it results in $|\mathcal{D}|_g = \sum_l|\mathcal{F}_l|_g$ and therefore, we may define 

\begin{equation}\label{eq:aD3_sum}
    a^3_{\mathcal{D}} = \sum_la_l^3 .
\end{equation}
By differentiating this relation twice with respect to the foliation time, one obtains,

\begin{equation}\label{eq:aD_sum}
    \frac{\ddot{a}_{\mathcal{D}}}{a_{\mathcal{D}}} = \sum_l\lambda_l\frac{\ddot{a}_l(t)}{a_l(t)}+\sum_{l\neq m}\lambda_l\lambda_m(H_l -H_m)^2 \, .
\end{equation}

\par Cosmological parameters can be defined for specific regions of interest, similarly to how they are defined in the standard Friedmann framework. These parameters are derived by dividing the Hamilton constraint (\autoref{eq:aD_dot}) by \(3H_{\mathcal{D}}^2\). The regions considered are denoted as $\mathcal{D}$ for the global domain, $\mathcal{M}$ for the overdense subregion, and $\mathcal{E}$ for the underdense subregion, with the generic label $\mathcal{F}$ used to represent any of these domains. The following parameters are defined:
\begin{widetext}
\begin{equation}
\Omega_{\mathcal{F}}^m := \frac{8\pi G}{3H_{\mathcal{D}}^2} \langle \rho \rangle_{\mathcal{F}}; \hspace{0.5 cm} \Omega_{\mathcal{F}}^\Lambda := \frac{\Lambda}{3H_{\mathcal{D}}^2};\hspace{0.5 cm}\Omega_{\mathcal{F}}^R := -\frac{\langle R \rangle_{\mathcal{F}}}{6H_{\mathcal{D}}^2};\hspace{0.5 cm}\Omega_{\mathcal{F}}^Q := -\frac{Q_{\mathcal{F}}}{6H_{\mathcal{D}}^2}.
\end{equation}
\end{widetext}
Applying these definitions, the Hamiltonian constraint (\autoref{eq:aD_dot}) for any domain $\mathcal{F}$ can be expressed as:

\begin{equation}
\Omega_{\mathcal{F}}^m + \Omega_{\mathcal{F}}^\Lambda + \Omega_{\mathcal{F}}^R + \Omega_{\mathcal{F}}^Q = \frac{H_{\mathcal{F}}^2}{H_{\mathcal{D}}^2}.
\end{equation}
This equation indicates that the sum of the dimensionless parameters equals 1, specifically for the global domain $\mathcal{D}$. However, in other regions, the sum can deviate from 1, depending on whether the expansion rate of the region $\mathcal{F}$ is greater or less than that of the region $\mathcal{D}$. In this work, we consider $\Lambda = 0$.

\section {Closing the system with scaling solutions}\label{sec:scale}

In the preceding section, we described the equations that linked various subregions within a partitioned space to the overall domain. In this section, we will focus on the most straightforward partitioning technique \cite{Wiltshire_2007,wiltshire_2007_1, Rasanen_2004,  Koksbang_2019, Koksbang2, Koksbang_PRL, Rasanen_2008, bose, Bose2013, Ali_2017, Pandey_2022, Pandey2023, Koksbang5, Koksbang6, Koksbang7, Koksbang8,  Koksbang2023}. As mentioned earlier,  the global domain is partitioned into overdense and underdense regions, labelled $\mathcal{M}$ and $\mathcal{E}$, respectively. 
In (\autoref{sec:buchert}), the averaged equations (\autoref{eq:aD_ddot})-(\autoref{eq:continuity}) do not form a closed system. One way to close the system is by imposing a specific equation of state, similar to what is done in Friedmannian models. Using the formalism of \cite{Weigand_et_al},  we can assume the following scaling laws for the backreaction and curvature terms, given by

\begin{equation}
Q_{\mathcal{F}} = Q_{\mathcal{F}_i} \, a_{\mathcal{F}}^n \quad \text{and} \quad \langle {\mathcal{R}} \rangle_{\mathcal{F}} =  {\mathcal{R}}_{\mathcal{F}_i} \, a_{\mathcal{F}}^p, \label{eq:scaling_ansatz}
\end{equation}
where $\mathcal{F}$  represents $\mathcal{M}$ or $\mathcal{E}$, and the subscript \(i\) indicates the initial state of the domain $\mathcal{F}$. The integrability condition (\autoref{eq:integrability}) links \(Q_{\mathcal{F}}\) and \(\langle R \rangle_{\mathcal{F}}\). On imposing (\autoref{eq:integrability}) on (\autoref{eq:scaling_ansatz}), two types of solution emerge: one with \(n = -6\) and \(p = -2\), and another where \(n = p\). The first solution is less interesting as it lacks the coupling between backreaction and curvature. The second solution, with \(n = p\), implies:

\begin{equation}
Q_{\mathcal{F}} = r_{\mathcal{F}} \langle \mathcal{R} \rangle_{\mathcal{F}} = r_{\mathcal{F}} \mathcal{R}_{\mathcal{F}_i} \, a_{\mathcal{F}}^n, \label{eq:curvature_backreaction_relation}
\end{equation}
where \(r_{\mathcal{F}}\) is determined by (\autoref{eq:integrability}) as:

\begin{equation}
r_{\mathcal{F}} = -\frac{n + 2}{n + 6}. \label{eq:r_factor}
\end{equation}

In the analysis presented in \cite{Weigand_et_al}, the value of the variable $n = -1$, as previously outlined in the work \cite{Li_et_al_2007},  does not replicate the history of structure formation from N-body simulations \cite{Weigand_et_al}. In the 
present study, we aim to select values of $n$ for our overdense $\mathcal{M}$ and underdense $\mathcal{E}$ consistent with the N-body structure formation simulations. In \cite{Weigand_et_al}, the simulation of the formation of the structure informed the selection of the equation of state, with a fixed value $\Omega_{\mathcal{D}_0}^{m}$ (the present time value of mass density parameter for the global domain $\mathcal{D}$) of $0.27$. Here we opt to constrain $\Omega_{\mathcal{D}_0}^{m}$ using observational data with new scaling solutions.

\par Within the above framework, distinct scaling factors are allocated to the regions of underdensity and overdensity. In this context, we designate the scaling term for the underdense region as $n_1$, while the scaling factor for the overdense region is represented as $n_2$. We select $n_2$, the scaling factor for the overdense region, to be $-2$, similar to a Friedmann scaling solution for simplicity. However, it is not feasible to set $n_1$ equal to $-2$, since this doesn't reflect the history of structure formation in the analysis, as indicated by the N-body simulation data \cite{simulation}. Consequently, we modify the scaling $n_1$ for the underdense region over the range from $-2$ to $-1$. Here, $-2$ represents Friedmann-like behaviour in terms of curvature $\mathcal{R}$, which can also be referred to as the first-order perturbation \cite{Li_et_al_2007}, while the scaling term $-1$ corresponds to the second-order perturbation \cite{Li_et_al_2007}. The motivation for modifying $n_1$ instead of $n_2$ is based on the following consideration. Though initially, both the overdense and underdense regions are set to contribute equally to the overall volume, as these regions undergo dynamic evolution over time, it is the underdense region that increasingly comes to dominate the total volume of the universe. Currently, the dominance of the underdense region is pronounced ($\sim 91\%$ \cite{Weigand_et_al}), playing an overwhelming role in structuring the universe, as structure formation becomes more prominent at the later stages of the universe's evolution. 
For each selection of $n_1$ within the interval $-2 \leq n_1 \leq -1$, we determine our parameter $\Omega^m_{\mathcal{D}_0}$ by performing the MCMC analysis \cite{mcmc1,mcmc2} using the Union 2.1 type Ia supernova data \cite{union}, to align with the structure formation history. Further details are provided in (\autoref{sec:ob}).  

\par For the overdense region denoted by $\mathcal{M}$, where the parameter $n_2$ is set to the value of $-2$, a curvature similar to that of the Friedmann model is evident and, notably, there is an absence of backreaction, indicated by $\Omega_{\mathcal{D}_0}^{\mathcal{Q}} = 0$. In contrast, in the underdense region denoted by $\mathcal{E}$, where $n_1$ differs from $-2$, a more complex interaction is observed between backreaction and curvature, indicating an interconnected relationship between these two elements. Backreaction in low-density regions arises due to differences in how various parts of these regions expand. Although underdense regions contain much less matter than overdense ones, they are not entirely empty, they still have small but finite matter densities. Within these regions, sub-domains with less matter expand at a different rate than those with more matter. The uneven distribution of matter leads to variations in the local expansion rate. When averaged over the entire underdense region, these variations contribute additional terms to the effective cosmological dynamics, resulting in a nonzero backreaction. 

\par In order to obtain the Hubble evolution of the global domain \({\mathcal{D}}\), it is necessary to solve equation (\autoref{eq:aD_dot}), which can be simplified to the form below when considering the scaling solution $a_{\mathcal{F}}^n$ for our subdomain $\mathcal{M}$ and $\mathcal{E}$:

\begin{equation}
H_{\mathcal{D}_0}^2 \left[ \Omega_{\mathcal{F}_0}^m \left(\frac{a_{\mathcal{F}_0}}{a_{\mathcal{F}}}\right)^3 + \Omega_{\mathcal{F}_0}^{RQ} \left(\frac{a_{\mathcal{F}_0}}{a_{\mathcal{F}}}\right)^{-n} \right] = \left(\frac{\dot{a}_{\mathcal{F}}}{a_{\mathcal{F}}}\right)^2, \label{eq:domain_evolution}
\end{equation}
where we define \(\Omega_{\mathcal{F}_0}^{RQ} := \Omega_{\mathcal{F}_0}^R + \Omega_{\mathcal{F}_0}^Q\). Instead of using \(a_{\mathcal{M}_0}\) and \(a_{\mathcal{E}_0}\) as free parameters, we can work with \(\lambda_{\mathcal{M}_0}\) and \(a_{\mathcal{D}_0}\). Note that the equation (\autoref{eq:domain_evolution}) holds even when $n_2=-2$ for the case of $\mathcal{M}$ subregion, representing vanishing backreaction. As a result, for the $\mathcal{M}$ domain we obtain $\Omega_{\mathcal{M}_0}^{RQ} = \Omega_{\mathcal{M}_0}^{R}$. To further constrain the parameters \(H_{\mathcal{D}_0}^2\), \(\Omega_{\mathcal{F}_0}^m\), \(\lambda_{\mathcal{M}_0}\), \(a_{\mathcal{D}_0}\), and \(\Omega_{\mathcal{F}_0}^{RQ}\)\, we assume a  Gaussian profile for the density fluctuations in the early universe (around \(z = 1000\)) \cite{Planck_2018}. This implies that \(\lambda_{\mathcal{M}i} \approx 0.5\) and \(a_{\mathcal{M}_i} = a_{\mathcal{E}_i} = \sqrt[3]{1/2}\) \cite{Weigand_et_al}. Given that the early universe was nearly homogeneous, we have \(\langle \rho \rangle_{\mathcal{D}_i} \approx \langle \rho \rangle_{\mathcal{M}_i} \approx \langle \rho \rangle_{\mathcal{E}_i}\). It follows that,

\begin{equation}
\Omega_{\mathcal{F}_0}^m \approx \lambda_{\mathcal{M}_i} \left(\frac{a_{\mathcal{D}_0}}{a_{\mathcal{F}_0}}\right)^3 \Omega_{\mathcal{D}_0}^m, \label{eq:matter_fraction}
\end{equation}
which allows us to simplify (\autoref{eq:domain_evolution}) by replacing \(\Omega_{\mathcal{M}_0}^m\) and \(\Omega_{\mathcal{E}_0}^m\) with \(\Omega_{\mathcal{D}_0}^m\). To further reduce the number of unknown parameters, we use (\autoref{eq:domain_evolution}) for both overdense $\mathcal{M}$ and underdense $\mathcal{E}$ regions at the present time, which yields

\begin{equation}
\frac{\lambda_{\mathcal{M}_i}}{ \lambda_{\mathcal{M}_0}} \Omega_{\mathcal{D}_0}^m + \Omega_{\mathcal{M}_0}^{RQ} = \frac{H_{\mathcal{M}_0}^2}{H_{\mathcal{D}_0}^2}, \label{eq:today_m}
\end{equation}

\begin{equation}
\frac{\lambda_{\mathcal{M}i}}{(1 - \lambda_{\mathcal{M}_0})} \Omega_{\mathcal{D}_0}^m + \Omega_{\mathcal{E}_0}^{RQ} = \frac{H_{\mathcal{E}_0}^2}{H_{\mathcal{D}_0}^2}, \label{eq:today_e}
\end{equation}

The above two equations, (\autoref{eq:today_m}) and (\autoref{eq:today_e}) together with the relation $H_{\mathcal{D}_0} = \lambda_{\mathcal{M}_0}H_{\mathcal{M}_0}\,+(1 - \lambda_{\mathcal{M}_0})H_{\mathcal{E}_0}$ which can be derived from (\autoref{eq:averaging}) can be used to eliminate \(\Omega_{\mathcal{M}_0}^{RQ}\). Another consistency condition allows us to eliminate \(\Omega_{\mathcal{E}_0}^{RQ}\), fixing the model without requiring detailed knowledge of the backreaction or curvature terms. The argument proceeds as follows. Consider the integral of (\autoref{eq:domain_evolution}),

\begin{equation}
H_{D_0}\int_0^t \mathrm{d}t' = \int_{\sqrt[3]{1/2}}^{a_{\mathcal{F}}} \left[ \frac{\Omega_{\mathcal{D}_0}^m}{2} \left(\frac{a_{\mathcal{D}_0}^{3}}{a'_{\mathcal{F}}}\right) + \Omega_{\mathcal{F}_0}^{\mathrm{RQ}} a_{\mathcal{F}_0}^{-n} \, {a'_{\mathcal{F}}}^{n+2} \right]^{-\frac{1}{2}} \mathrm{d}a'_{\mathcal{F}}.
\label{eq:time_integral}
\end{equation}
which leads to two functions $t_{\mathcal{M}}\left(a_{\mathcal{M}}\right)$ and $t_{\mathcal{E}}\left(a_{\mathcal{E}}\right)$ that depend on the integral's parameters.  Assuming the same time of the foliation in both the equations, $t_{\mathcal{M}_{0}} = t_{\mathcal{E}_{0}}$, (here we don't consider any
possible lapse between times in different regions \cite{wiltshire, Wiltshire_2007}),  we derive another relationship linking $\Omega^{\mathcal{R} \mathcal{Q}}_{\mathcal{M}_{0}}$ and $\Omega^{\mathcal{R} \mathcal{Q}}_{\mathcal{E}_{0}}$ with the other parameters. This enables the elimination of $\Omega^{\mathcal{R} \mathcal{Q}}_{\mathcal{E}_{0}}$, reducing the set to four free parameters: $H_{\mathcal{D}_{0}}$, $\Omega^{m}_{\mathcal{D}_{0}}$, $a_{\mathcal{D}_{0}}$, and $\lambda_{\mathcal{M}_{0}}$. We set $a_{\mathcal{D}_0}=1000$, following the normalization in \cite{Weigand_et_al} so that the recombination epoch ($z \approx 1000$) corresponds to $a_{\mathcal D} \approx 1$; this choice leaves the physics and resulting observables unchanged, as only ratios of $a_{\mathcal D}$ at different times enter the equations. We also set $|\lambda_{\mathcal{M}_0}| = 0.09$, with $H_{\mathcal{D}_0} = 100\hspace{0.05 cm}h$ km\,s$^{-1}$\,Mpc$^{-1}$. Subsequently, we constrain  $\Omega_{D_0}^m$ using observational data from the Union 2.1 Supernova Ia Data set \cite{union} as elaborated in (\autoref{sec:ob}).

\section {Analysis of $T_{21}$ Signal  within the Inhomogeneous Framework}\label{sec:par_plot}

   \begin{figure*}{H}
    \centering
    \begin{tabular}{cc}
	\includegraphics[trim={0 0 30 0},clip, width=0.5\textwidth]{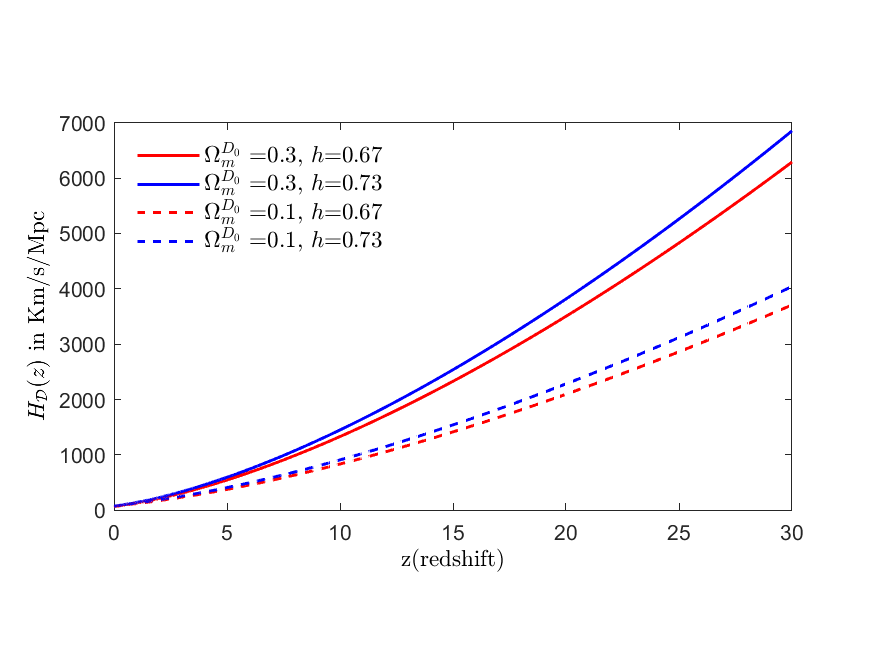}&
	\includegraphics[trim={0 0 30 0},clip, width=0.5\textwidth]{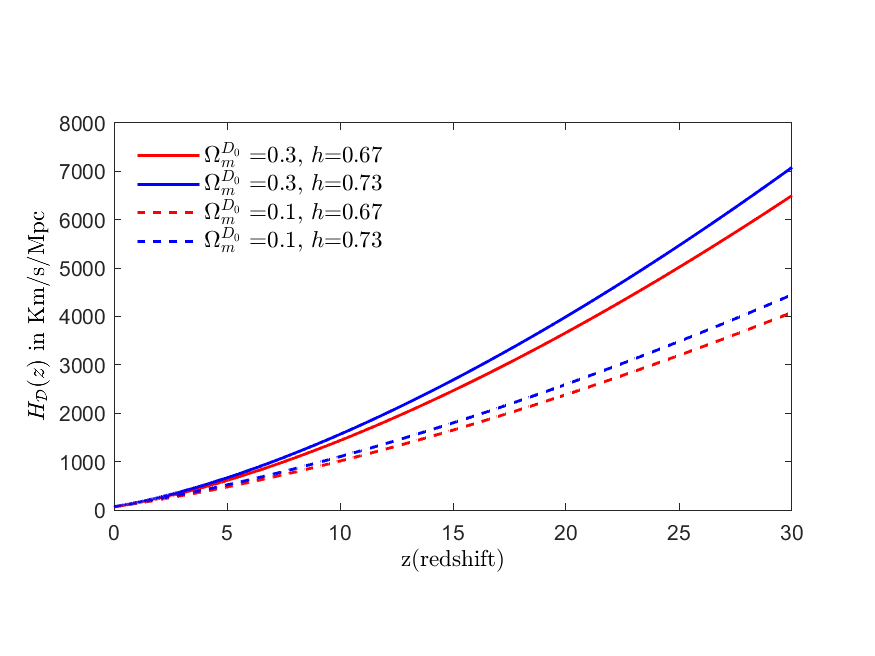}\\
	(a)&(c)\\
	\includegraphics[trim={0 0 40 0},clip, width=0.5\textwidth]{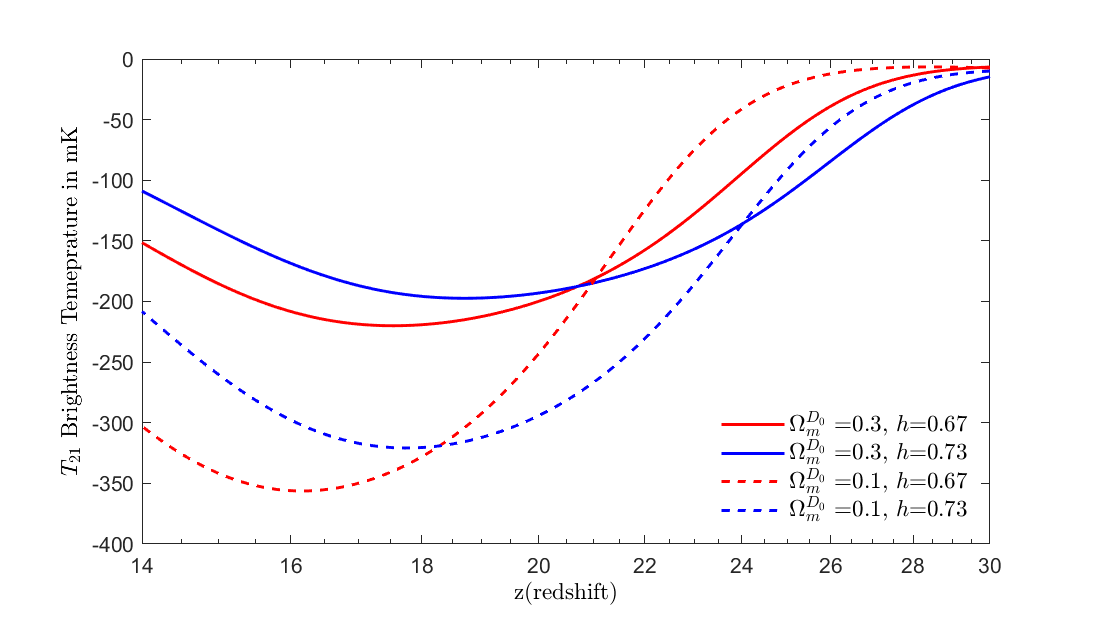}&
	\includegraphics[trim={0 0 40 0},clip, width=0.5\textwidth]{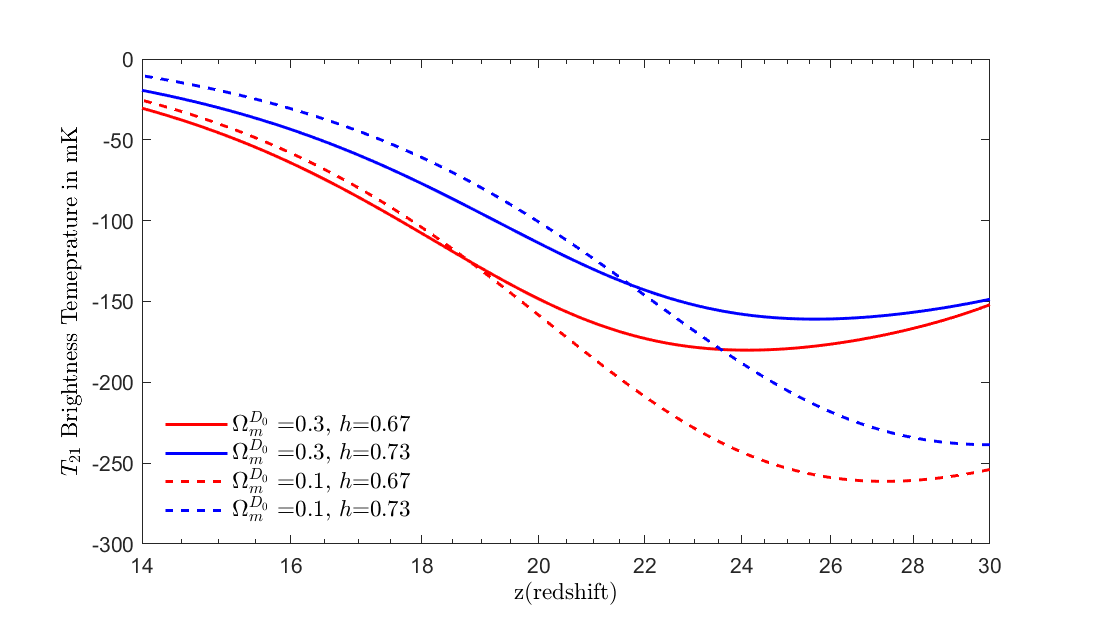}\\
	(b)&(d)\\
    \end{tabular}
 
   \caption{Hubble parameter and 21 cm brightness temperature evolution. \textbf{(a)} Hubble parameter $H_{\mathcal{D}}(z)$ versus redshift, $z$ for model parameters with $n_1 = -1.5$. \textbf{(b)} Corresponding $T_{21}$ plot for $n_1 = -1.5$. \textbf{(c)} Hubble parameter for $n_1 = -2$. \textbf{(d)} Corresponding $T_{21}$ plot for $n_1 = -2$. These panels demonstrate how variations in $\Omega_{m}^{D_0}$ and $h$ influence the optical depth and the $\mathrm{Ly\alpha}$ coupling, thereby shaping the absorption feature in $T_{21}$.}
      \label{fig:H_T_21}
\end{figure*}

\par We now study the behaviour of the brightness temperature $T_{21}$ for our model parameters  $\Omega^m_{\mathcal{D}_0}$, $h$ and $n_1$. Various choices of $\Omega^m_{\mathcal{D}_0}$, $h$, and $n_1$ give us different brightness temperature evolution, which we will analyze in this section. To
analyze the plots of (\autoref{fig:H_T_21}), let us focus on the brightness temperature expression given in (\autoref{eq:T21}), where we observe that the magnitude of $T_{21}$ depends on the optical depth. The spin temperature $T_s$ determines whether we get an absorption or an emission feature in the signal. We can obtain an absorption feature with respect to CMB if $T_s < \ T_{\gamma}$. The depth of the absorption feature is governed by $\tau$. An increase in the optical depth $\tau$ increases the magnitude of brightness temperature $T_{21}$ because $T_{21} \propto (1 - e^{-\tau})$. Further, if we decrease the spin temperature $T_s$ such that $T_s << \ T_{\gamma}$,  obtain  a dip in $T_{21}$. 

\par In subplot (a) of (\autoref{fig:H_T_21}), we plot the Hubble parameter as a function of redshift in the range $0 \leq z \leq 30$ where we choose $n_1 = -1.5$ for different sets of $\Omega^m_{\mathcal{D}_0}$ and $h$. Value of $H_{\mathcal{D}}(z)$ dictates the value of $\tau(z)$ as $\tau(z) \propto \frac{1}{H_{\mathcal{D}}(z)}$ which can be inferred from (\autoref{eq:tau}). Subplot (b) is the plot of brightness temperature in the redshift range $14 \leq z \leq 30$ for the same value of model parameters $\Omega^m_{\mathcal{D}_0}$, $h$ and $n_1(=-1.5)$ used in subplot (a) for the plot of $H_{\mathcal{D}}(z)$. We observe that at higher red-shifts and for a bigger value of $\Omega^m_{\mathcal{D}_0} = 0.3$,  the evolution of brightness temperature $T_{21}$ is dictated by the strength of $\mathrm{Ly\alpha}$ coupling where for higher value of $h$ the dip in the signal is more dominant. However, as $z$ decreases the domination is dictated by the $\tau$ part, {\it viz.}, $\tau \propto e^{-H}$ from  (\autoref{eq:T21}), and this causes more dip in $T_{21}$.  A lower value of $h$ enhances the absorption feature, which is also consistent with the subplot (a) of (\autoref{fig:H_T_21}), showing that the entire Hubble evolution is suppressed for lower $h$. For the low $\Omega^m_{\mathcal{D}_0}=0.1$ case,  we see a similar trend except a more prominent absorption feature of the brightness temperature of the 21 cm signal that lies in the EDGES range. This is due to the change in $H$, as $H$ has more suppressed evolution for low $\Omega^m_{\mathcal{D}_0}$, thus  lowering $\tau$,  which in turn amplifies the magnitude of $T_{21}$. In summary, a reduced expansion rate via lower $h$ or lower $\Omega^m_{\mathcal{D}_0}$ increases the 21 cm optical depth and thus amplifies the depth of the observed absorption feature.  

Similarly, in subplots (c) and (d), we plot the Hubble parameter and brightness temperature, by taking $n_1 = -2$. Since the scaling is chosen to be $-2$ for curvature,  by (\autoref{eq:curvature_backreaction_relation}) and (\autoref{eq:r_factor}), the backreaction vanishes. For this scaling choice, the subplot (d) of (\autoref{fig:H_T_21}) shows us that the brightness temperature for both values of $\Omega^m_{\mathcal{D}_0}$ displays a considerable amount of absorption dip at $z =30$. The evolution of $T_{21}$ signal is essentially controlled by Hubble evolution. A lower $\Omega^m_{\mathcal{D}_0}$ value suppresses $H_{\mathcal{D}}(z)$ as can be seen from the subplot (c) of (\autoref{fig:H_T_21}), which results in the increase of optical depth $\tau$. However, as the universe evolves with time, under the scaling $n_1 = -2$, the inter galactic medium quickly starts to heat by the $\mathrm{Ly\alpha}$ heating due to which $T_{21}$ evolution is no longer impacted much by Hubble evolution as $z$ decreases, and finally, all $T_{21}$ plots irrespective of the $\Omega^m_{\mathcal{D}_0}$ and $h$ values start converging to around $T_{21}\approx -30mk$.

 \par It may be noted that for both the scenarios ($n_1 = 1.5$ and $n_1 = -2$) discussed above, there is a particular stage (range of redshift) at which the evolution of the Hubble parameter predominantly influences the brightness temperature of the 21 cm signal. For $n_1 =-1.5$ it is around $z=16-22$ and for $n_1 = -2$ it is in the range $z=24-30$. In these regions, a notable decrease in brightness temperature is observed, particularly when the values of low $\Omega^m_{\mathcal{D}_0}$ are involved.

\section{Observational constraints}\label{sec:ob}

\begin{figure*}
    \centering
    \begin{tabular}{cc}
	\includegraphics[trim={0 5 0 0}, clip, width=0.5\textwidth]{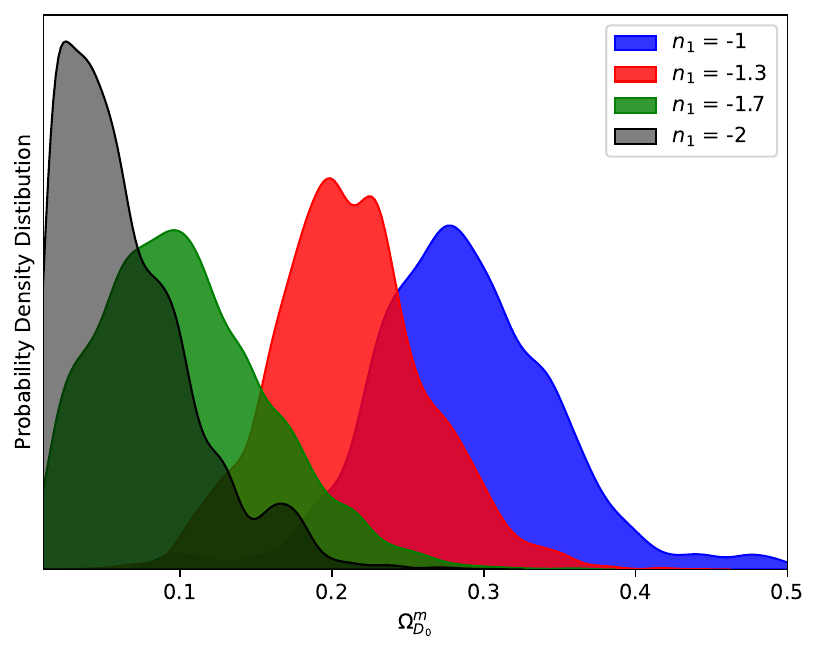}&
	\includegraphics[trim={25 5 30 5}, clip, width=0.55\textwidth]{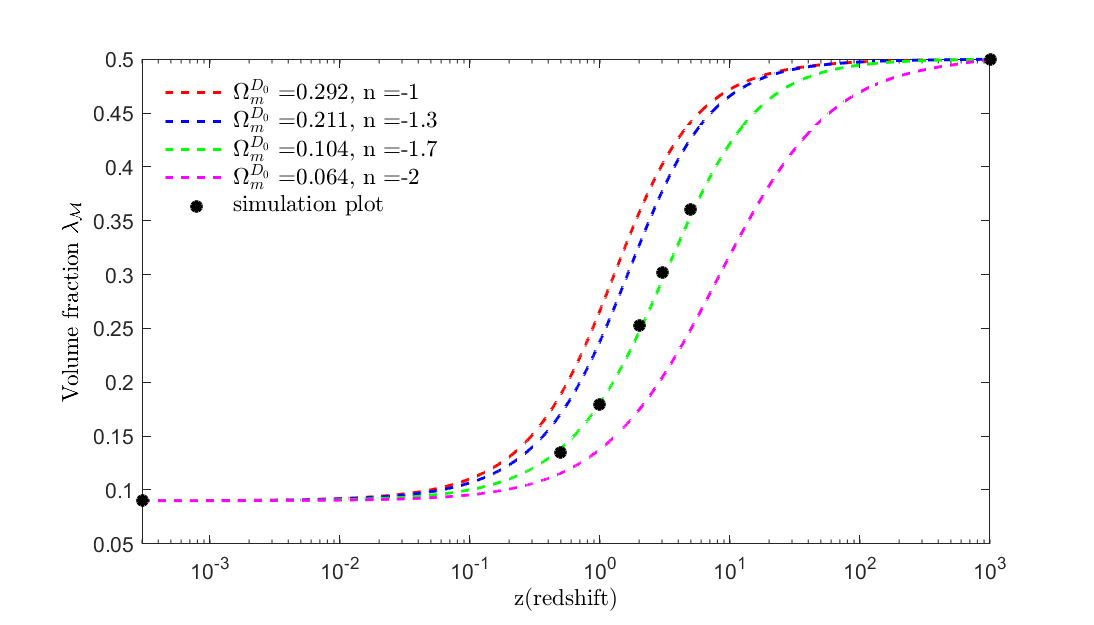}\\
	(a)&(b)\\
    \end{tabular}
 
   \caption{\textbf{(a)} Probability density distributions for the effective matter density parameter $\Omega_{m}^{D_0}$. The curves correspond to different choices of the scaling parameter $n_1$, as obtained from the MCMC analysis. The distributions illustrate the shift in the best fit $\Omega_{m}^{D_0}$ value with varying scaling behaviours for backreaction and curvature. \textbf{(b)} Evolution of the overdense region's volume fraction $\lambda_M = \frac{a_M^3}{a_D^3}$ versus redshift. The model predictions are compared with estimates extracted from N-body simulation data, showing consistency of the scaling solution with structure formation history.}
      \label{fig:omega_m_vol_frac}
\end{figure*}

     
     

\par In this section, we confront our model parameters against observation results of the Union 2.1 supernova Ia data \cite{union} to obtain the best-fit parameter values for our model. To align theoretically derived quantities from our model with the observational data (specifically redshift and angular diameter distance), we employ the covariant framework detailed in \cite{rasanen1,rasanen2}:

\begin{align}
    1+z &= \frac{1}{a_\mathcal{D}}\label{eq:covariant_sch_1}\\
    H_\mathcal{D}\frac{d}{dz}\left((1+z)^2H_\mathcal{D}\frac{dD_A}{dz}\right) &= -4\pi G\langle\rho\rangle_\mathcal{D} D_A.\label{eq:covariant_sch_2}
\end{align}
The foundational equation within the covariant framework (\autoref{eq:covariant_sch_1}) establishes a connection between the model-derived value $a_{\mathcal{D}}$ and the cosmological redshift $z$. Following this, the (\autoref{eq:covariant_sch_2}) links the model-computed quantity ${\langle \rho \rangle}_{\mathcal{D}}$ with the observable angular diameter distance $D_A$. The distance modulus can be calculated through standard cosmological distance relations using $D_A$, facilitating a comparison of our model with the Union 2.1 Supernova Ia data \cite{union}.

\par We constrain our model parameter $\Omega^m_{\mathcal{D}_0}$ using the Markov Chain Monte Carlo (MCMC) iteration method employing the \texttt{MCMCSTAT} package \cite{mcmc1,mcmc2}. We evaluate a total of $10,000$ events spanning the parameter range $\Omega^m_{\mathcal{D}_0} \in [0.01,0.5]$. We constrain $\Omega^m_{\mathcal{D}_0}$ for different values of $n_1$, which is the scaling solution for backreaction and curvature for the underdense region of our model. We choose $n_1$ $(-1,-1.3,-1.7,-2)$ between the range $-1$ and $-2$ where $-1$ corresponds to second order perturbation and $-2$ to first order perturbation \cite{Li_et_al_2007}. The values of $\Omega^m_{\mathcal{D}_0}$ with error levels values for different values of $n_1$ are tabulated in (\autoref{tab:1}). \\


 \begin{table}[!htbp]
  \centering
  \begin{tabular}{@{}cc@{}}
    \toprule
    $n_1$ & $\Omega^{m}_{\mathcal{D}_0}$ \\
    \midrule
    $-1$   & $0.292^{+0.054}_{-0.055}$ \\
    $-1.3$ & $0.211^{+0.049}_{-0.048}$ \\
    $-1.7$ & $0.104\pm0.056$         \\
    $-2$   & $0.064^{+0.040}_{-0.041}$ \\
    \bottomrule
  \end{tabular}
  \caption{Best‑fit values of the effective matter density parameter $\Omega^{m}_{D_0}$  
           for various choices of the scaling parameter $n_1$, as determined from the  
           MCMC analysis using the Union 2.1 Supernova Ia data.}
  \label{tab:1}
\end{table}

We find that our model suggests a low value of $\Omega^m_{\mathcal{D}_0}$ compared to the $\Lambda$CDM class of models. It may be noted, however, that some recent studies indicate that alternative cosmological models may favour a lower effective matter density than that predicted by $\Lambda$CDM. An independent measurement of $\Omega_m$ using gamma ray attenuation data has yielded  $\Omega_m =0.14^{+0.06}_{-0.07}$ \cite{Dom_nguez_2019}. Moreover, a universe comprised entirely of baryonic matter naturally leads to a critical density significantly lower than the standard model \cite{Gupta_2022, Gupta_2024}. Such complementary findings lend credence to our model's low effective $\Omega^m_{\mathcal{D}_0}$ value.

\begin{figure}[!htbp]
    \centering
    \includegraphics[trim={20 10 40 10}, clip, width=\columnwidth, keepaspectratio]{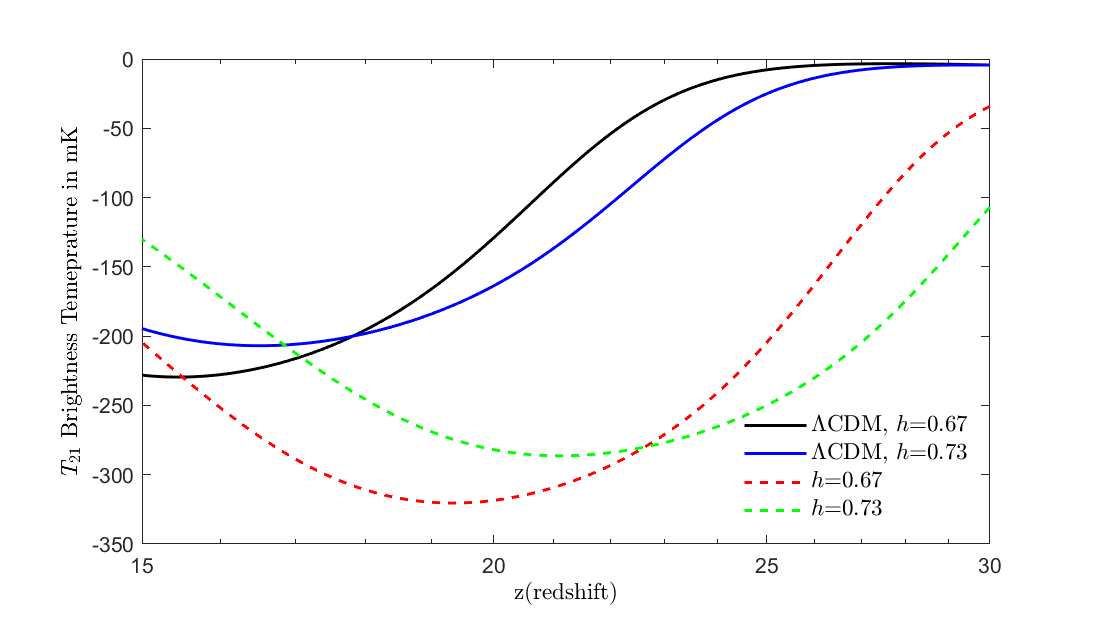}
   \caption{21 cm brightness temperature $T_{21}$ as a function of redshift in the range $15 \leq z \leq 30$ for different $H_0$ values. The plot shows that for $H_0 = 67\,$km/s/Mpc (PLANCK), the absorption dip in $T_{21}$ is deeper and more consistent with EDGES observations, compared to $H_0 = 73\,$km/s/Mpc (SH0ES) for the Backreaction model.}
    \label{fig:T21_vs_z}
    \label{fig:T_21_HT}
\end{figure}

In (\autoref{fig:omega_m_vol_frac}) subplot (a), we plot the probability density distribution for $\Omega^m_{\mathcal{D}_0}$ for the $n_1$ values mentioned in (\autoref{tab:1}).
(\autoref{fig:omega_m_vol_frac}) subplot (b) is the plot of the volume fraction of the overdense region as a function of redshift $z$, where the volume fraction of the overdense region ${\lambda}_{\mathcal{M}} = \frac{a^3_\mathcal{M}}{a^3_\mathcal{D}}$ is defined as the ratio of the volume occupied by the overdense region to the total volume of the global domain. The plot juxtaposes the volume fraction obtained from the N-body simulation (represented by black dots) with that obtained from our model. The data points are extracted from an analysis of an N-body simulation \cite{simulation} using a simple block separation technique described in \cite{Weigand_et_al}. The simulation box is segmented into a uniform grid to calculate $\lambda_\mathcal{M}$, which denotes the volume fraction of the overdense region, and the data points within each cell are enumerated. The volume of the densest cells is then aggregated until their cumulative number of points equals half of the total points in the simulation volume. This counting method is employed, assuming that the early universe is approximately homogeneous with  Gaussian perturbations. A grid length of $5h^{-1}Mpc$ is chosen to prevent overlap between the overdense and underdense regions. (Further details can be found in Appendix B-1 of \cite{Weigand_et_al}). From the volume fraction plot, we observe that for the scaling $n_1 = -1.7$, the plot matches consistently with the simulation data points. The value of $\Omega^m_{\mathcal{D}_0}$ obtained for $n_1 = -1.7$ from the MCMC analysis gives us $0.104\pm0.056$. The other scaling choices do not agree well with the simulation data.

\par Finally, we present a comparative analysis of the evolution of the
 $T_{21}$ signal between our backreaction model and the $\Lambda CDM$ model. We plot the brightness temperature $T_{21}$ as a function of redshift in the range $15 \leq z \leq 30$ for both models in (\autoref{fig:T_21_HT}). For our backreaction model we choose $n_1 = -1.7$ and $\Omega^m_{\mathcal{D}_0} = 0.104\pm0.056$ which are obtained by constraining these model parameters with the Union Supernova 2.1 Ia data \cite{union} and structure formation simulation data \cite{simulation}, and for the $\Lambda CDM$ model,  we use $\Omega_m=0.3$ \cite{union}. The comparison highlights the impact of varying the Hubble parameter $h$. The solid black and blue curves show the $\Lambda$CDM predictions for $h = 0.67$ and $h = 0.73$, respectively. These curves differ by no more than a few mk across the entire redshift range, indicating that in standard cosmology the 21\,cm absorption signal is largely insensitive to the value of $h$.

In contrast, the dashed red ($h = 0.67$) and dashed green ($h = 0.73$) curves correspond to our backreaction model in the Buchert framework. Here, a lower $h$ leads to a significantly deeper absorption trough, reaching approximately $-320$\,mK at $z \approx 18$, compared to about $-280$\,mK for the higher $h$ case. The redshift of the absorption minimum is also shifted by approximately $\Delta z \sim 3$. This occurs because reducing $h$ suppresses the overall Hubble expansion rate $H_{\mathcal{D}}(z)$, thereby increasing the 21\,cm optical depth $\tau \propto 1/H_{\mathcal{D}}(z)$ (\autoref{eq:tau}), which enhances the magnitude of the brightness temperature $T_{21}$. 
This strong $h$-dependence — entirely absent in the $\Lambda$CDM model —provides a clear observational signature of backreaction due to inhomogeneities. A precise measurement of the shape and depth of the global 21\,cm signal could thus offer a way to distinguish between standard cosmology and models with backreactiondriven expansion histories.

It may be noted that the prominent features of our analysis, {\it viz.}, the modified Hubble evolution and the $H_0$-dependent magnitude of the trough in the $T_{21}(z)$ signal are by no means an artefact of the simplified two-domain model used in the present analysis. As a robustness test of these
characteristic results shown above, we introduce a three-domain model and re-evaluate the 21-cm brightness temperature trough. The model and equations are 
presented in Appendix (\ref{app:three_domn}). A comparison of the two- and three-domain $T_{21}$ curves is displayed  in (\autoref{fig:T_21_HT_3D}). Though the exact depth and position of the trough does show  small differences depending on the chosen fractional volumes and scaling, our main conclusions remain intact.


\section{Conclusions}\label{sec:result}

 To summarize, in this work, we have elaborated and thoroughly analyzed an inhomogeneous cosmological model that strategically utilizes the 21 cm brightness temperature signal \cite{FURLANETTO2006181, BARKANA2016, Pritchard_2012} to explore one of the most significant challenges in contemporary cosmology: the Hubble tension \cite{Freedman_2021, Brout_2022}. Deviating from the conventional $\Lambda$CDM paradigm, our methodology uses Buchert's averaging formalism \cite{Buchert} to effectively represent the universe as an assemblage of regions with varying densities, specifically overdense and underdense regions \cite{Weigand_et_al, Wiltshire_2007}. By adopting this framework, we articulate scaling solutions for the backreaction and curvature components characterized by distinct evolutionary paths within these separate regions, thereby providing an alternate viewpoint on the dynamics of cosmic evolution.

\par Our model is carefully calibrated using observational data from the Union 2.1 Supernova Ia dataset \cite{union} and constraints from N-body simulation data on structure formation \cite{Weigand_et_al, simulation}. These calibrations yield best-fit values for $\Omega^m_{\mathcal{D}_0}$  
in conformity with our choice of scaling exponents for the backreaction and curvature terms. Moreover, the complex interplay between the optical depth, driven by the Hubble expansion, and the spin temperature, modulated by the $\mathrm{Ly\alpha}$ coupling mechanism, is crucial in determining the net absorption feature seen in the 21 cm line. By dissecting these contributions, we elucidate how variations in model parameters lead to marked differences in the brightness temperature evolution over the redshift range $14 \leq z \leq 30$, paving the way for a more nuanced interpretation of upcoming 21 cm observations and further exploration of inhomogeneous structures in the universe.

\par A central aspect of our study is the analysis of the 21 cm brightness temperature, $T_{21}$, which is intimately linked to both the optical depth and the spin temperature of neutral hydrogen. Our results demonstrate that the absorption feature observed in $T_{21}$ is highly sensitive to variations in the effective matter density parameter, $\Omega^m_{\mathcal{D}_0}$, and the Hubble parameter, $H_0$. Specifically, we show that lower values of $H_0$ (comparable to Planck-like measurements \cite{2020}) lead to a deeper absorption dip in the 21 cm signal, a result that is in better agreement with the EDGES observation \cite{Bowman}. This sensitivity underscores the potential of 21 cm cosmology as a robust diagnostic tool for probing the dynamics of the evolving universe and the underlying cosmological parameters. 

In conclusion, our work demonstrates that an inhomogeneous cosmological model with appropriately chosen scaling solutions offers a promising framework for exploring the Hubble tension further. The consistency of our model with both supernova data and structure formation simulations reinforces its viability as an alternative to conventional cosmological models. Looking ahead, further refinements of the model, such as through multiscale extensions \cite{T21_shas},  and the incorporation of additional observational datasets, such as those from future 21 cm experiments \cite{REACH, REACH2, DARE}, will be essential for fully unravelling the complexities of cosmic evolution and for resolving persistent discrepancies in the measurements of fundamental cosmological parameters.

\section{Acknowledgments}
SM would like to thank the Council of Scientific and Industrial Research (CSIR), Govt. of India, for funding through the CSIR-JRF-NET fellowship. SSP would like to thank the Council of Scientific and Industrial Research (CSIR), Govt of India, for funding through the CSIR-SRF-NET fellowship.

\appendix

\section{A model with three domains}
\label{app:three_domn}

Here we extend our two-domain backreaction model to a three-domain framework and re-evaluate the influence of the modified Hubble evolution on the $21$ cm brightness temperature trough.  In this context, we extend the division of our global domain $\mathcal{D}$ to encompass three subdomains: the underdense $\mathcal{E}$, the overdense $\mathcal{M}$, and an additional region, denoted as $\mathcal{A}$. This new subdomain, termed the ``ambient region," represents an intermediary segment of the universe that exists between clusters and voids. The introduction of the third domain $\mathcal{A}$ is well motivated by both simulations and observations of the cosmic web, which consistently reveal the existence of intermediate density structures such as filaments and sheets that bridge the gap between deep voids and dense clusters \cite{GalarragaEspinosa2020}. Including $\mathcal{A}$ enables us to quantify the influence of these environments on global backreaction and the resulting $21\,$ cm brightness temperature trough. A similar three domain model within the Buchert averaging framework was also investigated in \cite{Wiegand_thesis}. 

Similar to the approach used in the two-domain scenario, we partition the global domain $\mathcal{D}$ into specific subregions designated as $\mathcal{E}$, $\mathcal{M}$, and $\mathcal{A}$ for our analysis of a three-domain context. These subregions are generically represented by the symbol $\mathcal{F}$. This decomposition allows us to express the global domain $\mathcal{D}$ as $\mathcal{D}\,=\cup_{l}\mathcal{F}_{l}$, where $\mathcal{F}\,:=\cup_{\alpha}\mathcal{F}^{(\alpha)}_{l}$ and $\mathcal{F}^{(\alpha)}_{l}\cap\mathcal{F}^{(\beta)}_{m}=\phi$ for all $\alpha \neq \beta$ and $l \neq m$.

\par As elaborated on earlier in (\autoref{sec:buchert}), the series of equations outlined in (\autoref{eq:aD_ddot} - \autoref{eq:continuity}) does not constitute a closed system. To achieve closure for this system, we introduce the assumption of a scaling solution for both backreaction ($\mathcal{Q}_{\mathcal{F}}$) and curvature ($\mathcal{R}_{\mathcal{F}}$), as outlined in (\autoref{eq:scaling_ansatz}). Subsequently, we apply the integrability condition specified in (\autoref{eq:integrability}) to derive a relationship between $\mathcal{Q}_{\mathcal{F}}$ and $\mathcal{R}_{\mathcal{F}}$ expressed as follows:
\begin{equation}
    \mathcal{Q}_{\mathcal{F}}=r_{\mathcal{F}}\mathcal{R}_{\mathcal{F}} \label{QR_rel}
\end{equation}
with $r_{\mathcal{F}}=-\frac{n+2}{n+6}$.
 In a manner analogous to our analysis of the two-domain scenario, we postulate a unique scaling solution value, represented as $n$, for each discrete subdomain within our three-domain framework. These distinct values are labeled $n_1$, $n_2$, and $n_3$, corresponding specifically to the regions denoted by $\mathcal{E}$, $\mathcal{M}$, and $\mathcal{A}$, respectively.

\par In order to derive the Hubble evolution for the global domain $\mathcal{D}$ within our three-domain model, we apply (\autoref{eq:averaging}). This requires us to individually determine the Hubble evolution within each constituent subregion: $\mathcal{E}$, $\mathcal{M}$, and $\mathcal{A}$. Given the assumption of scaling resolution concerning backreaction and curvature for each subregion, the Hubble evolution can be obtained using the (\autoref{eq:domain_evolution}) for each subregion. Accordingly, the formulation of (\autoref{eq:domain_evolution}) for subregions $\mathcal{E}$, $\mathcal{M}$, and $\mathcal{A}$ is expressed as:
\begin{widetext}
\begin{align}
H_{\mathcal{D}_0}^2 \left[ \Omega_{\mathcal{E}_0}^m \left(\frac{a_{\mathcal{E}_0}}{a_{\mathcal{E}}}\right)^3 + \Omega_{\mathcal{E}_0}^{RQ} \left(\frac{a_{\mathcal{E}_0}}{a_{\mathcal{E}}}\right)^{-n_1} \right] = \left(\frac{\dot{a}_{\mathcal{E}}}{a_{\mathcal{E}}}\right)^2,
\label{eq:domain_evolution_E} \\
H_{\mathcal{D}_0}^2 \left[ \Omega_{\mathcal{M}_0}^m \left(\frac{a_{\mathcal{M}_0}}{a_{\mathcal{M}}}\right)^3 + \Omega_{\mathcal{M}_0}^{RQ} \left(\frac{a_{\mathcal{M}_0}}{a_{\mathcal{M}}}\right)^{-n_2} \right] = \left(\frac{\dot{a}_{\mathcal{M}}}{a_{\mathcal{M}}}\right)^2,
\label{eq:domain_evolution_M}\\
H_{\mathcal{D}_0}^2 \left[ \Omega_{\mathcal{A}_0}^m \left(\frac{a_{\mathcal{A}_0}}{a_{\mathcal{A}}}\right)^3 + \Omega_{\mathcal{A}_0}^{RQ} \left(\frac{a_{\mathcal{A}_0}}{a_{\mathcal{A}}}\right)^{-n_3} \right] = \left(\frac{\dot{a}_{\mathcal{A}}}{a_{\mathcal{A}}}\right)^2.
\label{eq:domain_evolution_A}
\end{align}
\end{widetext}

\par The matter density parameter for each region is determined from the present day density contrasts $\delta_{\mathcal{M}_0}$ and $\delta_{\mathcal{E}_0}$, which define the separation between the regions. They are given by:
\begin{widetext}
\begin{align}
\Omega_{\mathcal{E}_0}^m &= \Omega_{\mathcal{D}_0}^m\left(1+ \delta_{\mathcal{E}_0}\right), \label{eq:mass_frac_E}\\
\Omega_{\mathcal{M}_0}^m &= \Omega_{\mathcal{D}_0}^m\left(1+ \delta_{\mathcal{M}_0}\right), \label{eq:mass_frac_M}\\
\Omega_{\mathcal{A}_0}^m &= \Omega_{\mathcal{D}_0}^m
\frac{1 - \lambda_{\mathcal{E}}^0\left(1 + \delta_{\mathcal{E}}^0\right) - \lambda_{\mathcal{M}}^0\left(1 + \delta_{\mathcal{M}}^0\right)}{\lambda_{\mathcal{A}}^0}.
\end{align}
\end{widetext}
The density contrasts are defined as
\begin{eqnarray}
\delta_{\mathcal{E}_0} &=& \frac{\lambda_{\mathcal{E}}^i}{\lambda_{\mathcal{E}}^0} - 1, \label{eq:denscont_E} \\   
\delta_{\mathcal{M}_0} &=& \frac{\lambda_{\mathcal{M}}^i}{\lambda_{\mathcal{M}}^0} - 1. \label{eq:denscont_M}
\end{eqnarray}
Here $\Omega_{\mathcal{D}_0}^m$ is the mass density fraction of the global domain $\mathcal{D}$; $\lambda_{\mathcal{E}}^i$ and $\lambda_{\mathcal{M}}^i$ are the volume fractions of the overdense and 
underdense regions at the recombination epoch ($z \approx 1000$), and $\lambda_{\mathcal{E}}^0$ and $\lambda_{\mathcal{M}}^0$ are the corresponding fractions at the present epoch ($z = 0$). These relations follow from the standard definition $\delta = (\rho - \bar{\rho})/\bar{\rho}$ together with mass conservation for each domain, which implies that density scales inversely with the domain volume. Since at $z \approx 1000$ the densities of all regions are equal up to small Gaussian perturbations \cite{Weigand_et_al}, the present day density contrasts depend only on the relative change in volume fraction between the early and present epochs.

\par The quantities $\Omega_{\mathcal{E}_0}^{RQ}$, $\Omega_{\mathcal{M}_0}^{RQ}$, and $\Omega_{\mathcal{A}_0}^{RQ}$ are determined from a consistency condition that requires the proper-time function defined in (\autoref{eq:time_integral}), obtained by integrating (\autoref{eq:domain_evolution_E})--(\autoref{eq:domain_evolution_A}) for each region, to have the same value today. This requirement leads to the condition,
\begin{widetext}
\begin{equation}
t_1\left(a_{\mathcal{M}_0}, \Omega_{\mathcal{M}_0}^{RQ}, H_{\mathcal{D}_0}, 
\Omega_{\mathcal{D}_0}^m, a_{\mathcal{D}_0}, \lambda_{\mathcal{M}_0}, \lambda_{\mathcal{E}_0} \right) 
= t_2\left(a_{\mathcal{E}_0}, \Omega_{\mathcal{E}_0}^{RQ}, H_{\mathcal{D}_0}, 
\Omega_{\mathcal{D}_0}^m, a_{\mathcal{D}_0}, \lambda_{\mathcal{M}_0}, \lambda_{\mathcal{E}_0} \right) 
= t_3\left(a_{\mathcal{A}_0}, \Omega_{\mathcal{A}_0}^{RQ}, H_{\mathcal{D}_0}, 
\Omega_{\mathcal{D}_0}^m, a_{\mathcal{D}_0}, \lambda_{\mathcal{M}_0}, \lambda_{\mathcal{E}_0} \right),
\label{eq:time_eqlty}
\end{equation}
\end{widetext}
from which the parameters $\Omega_{\mathcal{E}_0}^{RQ}$, $\Omega_{\mathcal{M}_0}^{RQ}$, and $\Omega_{\mathcal{A}_0}^{RQ}$ can be eliminated.
\par We assign $\Omega_{\mathcal{D}_0}^m = 0.1$ and $a_{\mathcal{D}_0} = 1000$ for the three domain model. The value of $\Omega_{\mathcal{D}_0}^m$ is derived from constraints based on observations and simulations for the two-domain scenario (refer to \autoref{sec:ob}). In addition, for the three domain model we employ, we establish the scaling solutions to be $n_1 = -1.7$ for the underdense region and $n_2 = -2$ for the overdense region. The same scaling solutions apply to both underdense and overdense regions in our two-domain model, serving as best fit values derived from observational and simulation data (see \autoref{sec:ob}). For the ambient region $\mathcal{A}$, opting for a scaling factor $n_3$ in the range of $-2$ to $-1.7$ is a reasonable choice. Accordingly, in this analysis, we have selected $n_3=-1.8$ as the assumed value.

\begin{figure}[!htbp]
    \centering
    \includegraphics[trim={20 10 40 10}, clip, width=\columnwidth, keepaspectratio]{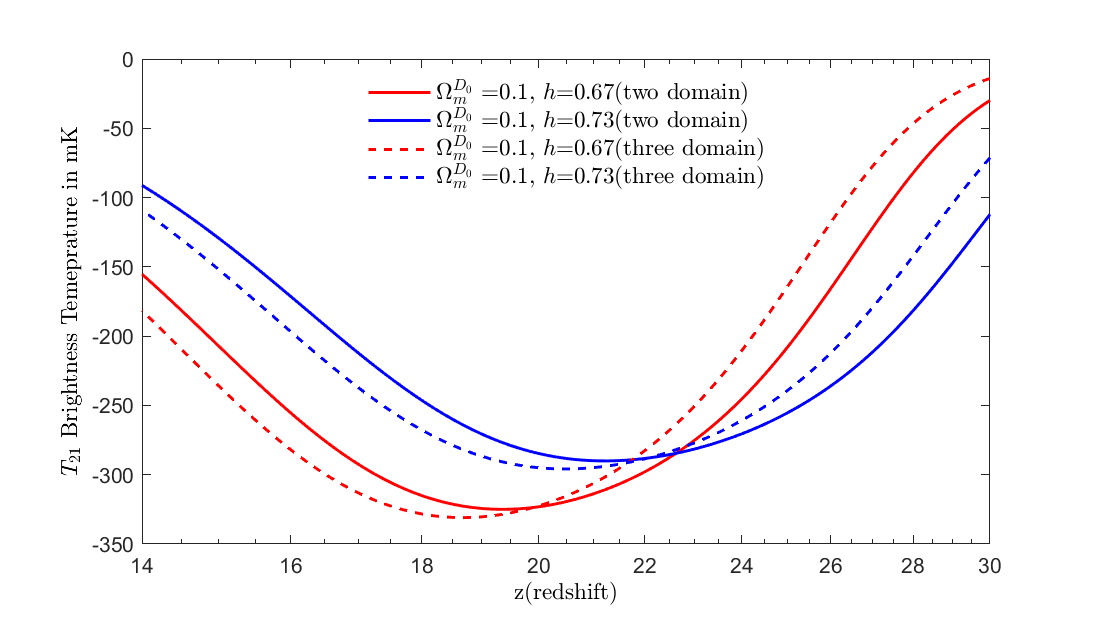}
   \caption{The 21-cm brightness temperature $T_{21}$ for $14 \leq z \leq 30$ is shown for two-domain (solid) and three-domain (dashed) backreaction models with $\Omega_{m}^{D_0} = 0.1$. For both $h = 0.67$ (red) and $h = 0.73$ (blue), higher $h$ yields a shallower, earlier trough. The three-domain case slightly modifies the trough depth and  position.}
  \label{fig:T_21_HT_3D}
\end{figure}

\par We set the density contrasts ($\delta$) and volume fractions ($\lambda$) for the overdense and underdense regions using constraints from both numerical simulations and galaxy surveys. Large-scale simulations of the cosmic web \cite{Cautun:2014rha,AragonCalvo:2010,libeskind2018,Einasto_2021,Romero2009} consistently find that voids occupy $\sim70$--$80\%$ of the volume, with walls and filaments accounting for most of the remainder, and compact cluster cores contributing less than $1\%$. Observational analysis from SDSS  report void filling factors of $\sim60$--$65\%$ \cite{Sutter_2012,Pan_2012} and overdense clusters taking about $1\%$ \cite{Einasto_2018,Einasto2019} of volume fractions, and recent DESI void catalogs \cite{DESI_2025} show comparable results. The mean void densities are typically $\sim0.1$--$0.2\,\bar{\rho}$ (corresponding to $\delta\simeq-0.8$ to $-0.9$), while the overdense knots reach overdensities several times the mean, ranging from $\delta\approx4$ to $\delta\approx10$ \cite{Einasto_2018,Einasto_2021,Romero2009,libeskind2018}. Following the web classification analysis of Libeskind et al. \cite{libeskind2018}, we adopt fiducial present-day values $\delta_{\mathcal{E}}^0=-0.8$ and $\lambda_{\mathcal{E}}^0=0.70$ for underdense regions, and $\delta_{\mathcal{M}}^0=9$ and $\lambda_{\mathcal{M}}^0=0.01$ for overdense regions. We now investigate how the generalization to a three domain affects the resulting 21-cm brightness temperature signal.

\par The (\autoref{fig:T_21_HT_3D})  compares the 21-cm brightness temperature $T_{21}(z)$ for our original two-domain backreaction model (solid lines) and the extended three-domain model (dashed lines), with the global matter fraction fixed at $\Omega_{m}^{D_0} = 0.1$. Red curves correspond to $h = 0.67$ and blue to $h = 0.73$. Both the two-domain and three-domain models produce a deeper minimum when $h$ is lower(0.67) (red curves are deeper than blue for corresponding model types). The three-domain construction slightly modifies the trough amplitude and shifts its precise redshift compared with the two-domain case, but it does not remove the absorption feature nor the  $H_0$-dependence; thus, the presence of a deep trough at lower $h$ is robust to this modification of the domain partitioning.

\bibliographystyle{apsrev4-2-author-truncate.bst}
\bibliography{ref.bib}

\end{document}